\documentclass[fleqn,usenatbib]{mnras}

\usepackage{newtxtext,newtxmath}
\usepackage[T1]{fontenc}

\usepackage{xspace}
\usepackage[export]{adjustbox}
\usepackage{float}
\usepackage{lineno}
\usepackage{gensymb}
\usepackage{graphicx}

\usepackage[hang]{footmisc}

\renewcommand{\arcmin}{\ensuremath{^\prime}\xspace}
\renewcommand{\arcsec}{\ensuremath{^{\prime\prime}}\xspace}

\newcommand{\SNe}{SNe Ia}

\defcitealias{hounsell_simulations_2018}{H18}
\defcitealias{troxel_synthetic_2021}{T21}
\def\HST{{\it HST}\xspace}

\title[Roman SN Image Sims]{A Synthetic Roman Space Telescope High-Latitude Time-Domain Survey:\\ Supernovae in the Deep Field}

\author[Wang et al.]{
Kevin X. Wang,$^{1}$
Dan Scolnic,$^{1}$
M. A. Troxel,$^{1}$
Steven A.~Rodney,$^{2}$
Brodie Popovic,$^{1}$
Caleb Duff,$^{2}$
\newauthor{
Alexei V. Filippenko,$^{3}$
Ryan J. Foley,$^{4}$
Rebekah Hounsell,$^{5,6}$
Saurabh W. Jha,$^{7}$
David O. Jones,$^{4,8}$
}
\newauthor{
Bhavin A.\ Joshi,$^{9}$
Heyang Long,$^{10}$
Phillip Macias,$^{4}$
Adam G. Riess,$^{9,11}$
Benjamin M. Rose,$^{1}$
Masaya Yamamoto$^{1}$
}
\\
$^{1}$Department of Physics, Duke University, Durham, NC, 27708, USA.\\
$^{2}$Department of Physics and Astronomy, University of South Carolina\\
$^{3}$Department of Astronomy, University of California, Berkeley, CA 94720-3411, USA\\
$^{4}$Department of Astronomy \& Astrophysics, University of California, Santa Cruz, CA 95064\\
$^{5}$University of Maryland, Baltimore County, Baltimore, MD 21250, USA\\
$^{6}$NASA Goddard Space Flight Center, Greenbelt, MD 20771, USA\\
$^{7}$Department of Physics and Astronomy, Rutgers, the State University of New Jersey, Piscataway, NJ 08854, USA\\
$^{8}$NASA Einstein Fellow\\
$^{9}$Department of Physics and Astronomy, Johns Hopkins University, Baltimore, MD 21218, USA\\
$^{10}$Department of Physics, The Ohio State University, 191 West Woodruff Avenue, Columbus, OH 43210, USA\\
$^{11}$Space Telescope Science Institute, Baltimore, MD 21218
}

\date{Accepted XXX. Received YYY; in original form ZZZ}

\pubyear{2015}

\begin{document}
\label{firstpage}
\pagerange{\pageref{firstpage}--\pageref{lastpage}}
\maketitle

\begin{abstract}
NASA will launch the \textit{Nancy Grace Roman Space Telescope} (\textit{Roman}) in the second half of this decade, which will allow for a generation-defining measurement of dark energy through multiple probes, including Type Ia supernovae (\SNe{}).  To improve decisions on survey strategy, we have created the first simulations of realistic \textit{Roman} images that include artificial \SNe{} injected as point sources in the images.  Our analysis combines work done on \textit{Roman} simulations for weak gravitational lensing studies as well as catalog-level simulations of SN~Ia samples. We have created a time series of images over two years containing $\sim 1,050$ \SNe{}, covering a 1 square degree subarea of a planned 5 square degree deep survey.  We have released these images publicly for community use along with input catalogs of all injected sources.  We create secondary products from these images by generating coadded images and demonstrating recovery of transient sources using image subtraction. We perform first-use analyses on these images in order to measure galaxy-detection efficiency, point source-detection efficiency, and host-galaxy association biases. The simulated images can be found here: {\url{https://roman.ipac.caltech.edu/sims/SN_Survey_Image_sim.html}}.
\end{abstract}


\section{\label{sec:level1}Introduction}

\bigskip

 The \textit{Nancy Grace Roman Space Telescope} (hereafter \textit{Roman}) is a NASA mission that will make advances into a wide range of astrophysics, and should be able to make an unprecedented measurement of the nature of dark energy with multiple probes \citep{spergel_wide-field_2015}.  The Project is currently in Key Design Phase C (final design and fabrication) and most of the specifications are now set.\footnote{\url{https://roman.gsfc.nasa.gov/science/WFI_technical.html}}  The 2.4\,m space-based telescope is optimized for observations in the near-infrared (NIR), with a Wide Field Instrument (WFI) that can accommodate eight filters spanning a wavelength range of 0.48--2.3\,$\mu$m and has a large field of view (FoV) of 0.281 square degrees.

Because of its location at L2, large aperture size, and FoV, \textit{Roman} will be extremely well-suited for studies of cosmological growth and expansion on large scales.  The extragalactic survey will be divided into two main parts, with a wide-area High-Latitude Survey (HLS) that will enable studies of weak lensing and galaxy clustering, as well as a smaller-area High-Latitude Time-Domain Survey.  As originally formulated in the \textit{Roman} Science Requirement Document,\footnote{\url{https://science.nasa.gov/science-pink/s3fs-public/atoms/files/Kruk_APAC_Oct2018_r2.pdf}} the top goal of the time-domain survey is to study the expansion history of the Universe from redshift $z\approx0.1$ to $z\approx2$.  Current samples of Type Ia supernovae (\SNe{}) used for cosmological analyses \citep{Pantheon+} contain on the order of 1,500 objects, though the number of \SNe{} with well-sampled light curves (cadence better than once every 10 rest-frame days per filter) above $z>1$ is only $\sim 20$.  \textit{Roman} has the ability to discover and measure over 10,000 \SNe{}, with a significant fraction of these at $z>1$ \citep{hounsell_simulations_2018}. In this work, we create the first simulated images of the time-domain survey to better prepare for the future mission.  
    
    Precision measurements of \SNe{} across this redshift range are critical for understanding the equation-of-state parameter for dark energy, $w = P/\rho$, where $P$ is the pressure and $\rho$ is the energy density. A common parametrization of $w$ combines a constant with a term that depends linearly on the scale factor ($a$), such that $w = w_{0} + w_{a}(1 - a)$ \citep{PhysRevLett.90.091301}. The recent Pantheon+ analysis of a composite sample containing $\sim1500$ \SNe{} found $w_0=-0.841^{+0.066}_{-0.061}$ and $w_a=-0.65^{+0.28}_{-0.32}$ (after including constraints from measurements of the cosmic microwave background and baryon acoustic oscillations), which shows that currently there is only a very weak constraint on the evolution of dark energy \citep{Pantheon+Analysis}.  While current systematic constraints are of comparable magnitude to the statistical constraints, both can be improved with a larger sample and redshift range.
    
Although at present there is not a single defined survey strategy for the SN key project in terms of cadence, depth, filter choices, etc., we rely here on the initial study done by \citet{hounsell_simulations_2018} (hereafter \citetalias{hounsell_simulations_2018}) that evaluated multiple survey strategies and performed catalog-level simulations with SNANA \citep{kessler_snana:_2009} to predict final constraints on dark-energy parameters from the programs in terms of the figure-of-merit (FoM), first introduced for dark-energy studies by the Dark Energy Task Force \citep{albrecht_report_2006}. For simplicity in this analysis, we choose a single, representative survey strategy from those presented by \citetalias{hounsell_simulations_2018}. With the image simulations presented here, we can measure detection efficiencies and other survey characteristics that can be used for the less computationally expensive catalog level simulations.
    
   To enable the image simulations, we rely on the work done by \citet{troxel_synthetic_2021} (hereafter \citetalias{troxel_synthetic_2021}) which created realistic simulated \textit{Roman} images that covered 6 sq. degrees following a reference survey design for the High-Latitude Imaging Survey (HLIS). They simulated multiple effects that can cause systematic bias in cosmological analysis, such as background light, Poisson noise, reciprocity failure, dark current, classical nonlinearity, and interpixel capacitance. These detector effects are described in detail by \citetalias{troxel_synthetic_2021} and are also summarized in Table~\ref{tab:detector_effects}. They include both stars and galaxies in their images.  In this study, we can use the same architecture for our simulated images where we overlay \SNe{} as stellar point sources on top of galaxy images and analyze time-series of images.

\begin{table*}
\begin{centering}
\caption{\label{tab:detector_effects} Detector effects; they are discussed in more detail by \citet{troxel_synthetic_2021}. GalSim is open-source software used to simulate images \citep{Rowe2015} and is discussed in Section 2.2.2.}
%\hspace{-.5in}
\begin{tabular}{p{0.2\textwidth}p{0.3\textwidth}p{0.425\textwidth}}
Effect & Description & Relevant functions within the GalSim module \\
\hline
Background Light & Zodiacal and stray light and thermal emission of the telescope & Uses roman.getSkyLevel\\
Photon count noise & Poisson noise on photon counts & Uses noise.PoissonNoise \\
Reciprocity Failure & Response to high flux is larger than to low flux & Uses roman.roman\_detectors.addReciprocityFailure  \\
Electron Quantization & Electron count is rounded to nearest integer &  Uses detectors.quantize \\
Dark Current & Poisson noise caused by thermal electrons & Uses DeviateNoise on PoissonDeviate \\
Classical Nonlinearity & Nonlinear charge-to-voltage response  & Uses detectors.applyNonlinearity with roman.NLfunc\\
Interpixel Capacitance & Pixels are affected by charge on neighboring pixels & Uses roman.applyIPC \\
Read Noise & Noise from electronics as charge is converted to digital values & Gaussian noise with standard deviation from roman.read\_noise \\
e to ADU & Electron counts are converted to ADUs & Currently applies gain of 1 since actual gain is not yet known \\
ADU quantization & ADU rounded to nearest integer & Uses detectors.quantize \\
\hline
\end{tabular}
\end{centering}    
\end{table*}

Artificial SN injection has been used in many studies \citep{Pain_1996, Sullivan06}, including by the Dark Energy Survey (DES) \citep{kessler_first_2019} and LSST DESC \citep{LSST_DC2}.  As discussed by \citet{kessler_first_2019},  DES used artificial sources to track failures of their image-processing pipeline, to characterize their efficiency, and finally to trace biases in recovered photometry all the way to biases in the recovered cosmology \citep{brout_first_2019}.  LSST DESC has similar goals with their simulated sky survey called ``Data Challenge 2" (DC2) \citep{LSST_DC2} and the artificial transients have been first analyzed by \citet{BrunoDC2}. In this paper, we utilize infrastructure for characterizing the relation between SN~Ia properties and host-galaxy properties added as part of the SNANA package \citep{popovic_assessment_2019}. 
  
 There are a number of time-domain analyses that can be done with a large image simulation, including tests of the impact of saturation on the bright end and recovery near the cores of galaxies.  We provide the code used to perform the analysis, which may be used as a base for more complicated tests.\footnote{\url{https://github.com/KevinXWang613/SNAnalysisCode}} One such example is the work in \cite{Rubin21}, which establishes a forward-modeling photometry pipeline to be used on Roman images.
    
The overview of this paper is as follows. In Section \ref{sec:methods}, we describe our methodology for constructing simulated images. Section \ref{sec:supernovae} presents the details of generating and including \SNe{} in images. We provide analyses in Section \ref{sec:analysis}, such as measuring how often a SN~Ia is likely to be mistakenly associated with the wrong host galaxy and measuring the SN~Ia detection efficiencies as a function of magnitude. The details about processing images in preparation for analysis, particularly for creating difference images and detecting point sources, are also in this section. In Section \ref{sec:conclusions}, we discuss the implications of this work for design of the \textit{Roman} SN~Ia cosmology survey, and describe future work.

\section{Methods}\label{sec:methods}

\begin{figure}
    \centering
    \includegraphics[width=\columnwidth]{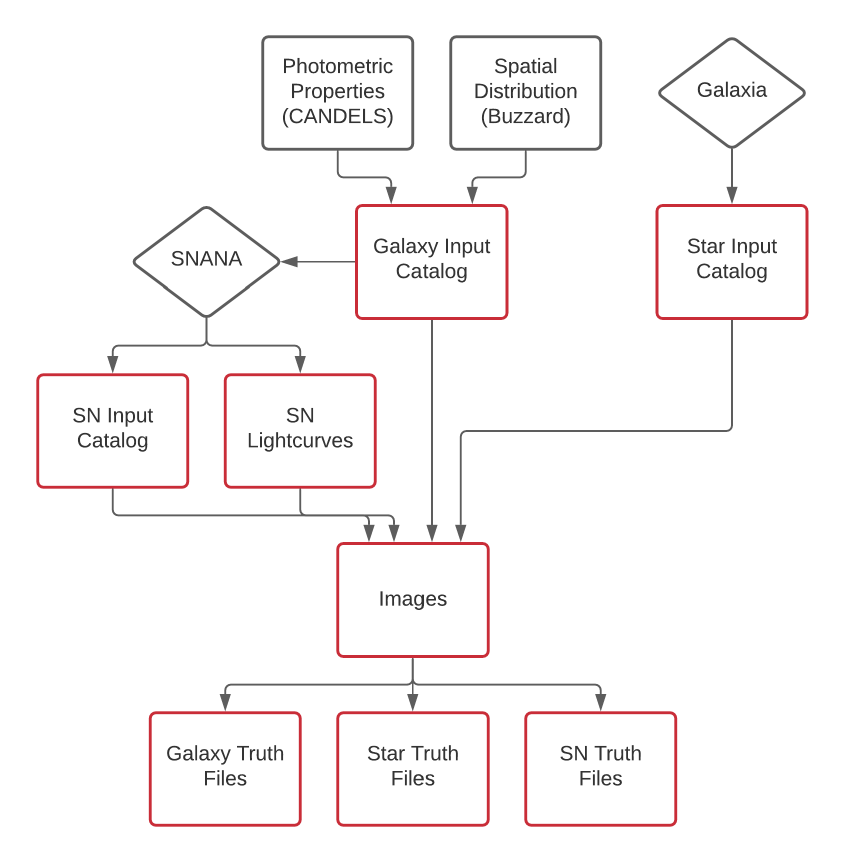}
    \caption{Flowchart depicting the process through which output products are created. Boxes labeled in red are the output files that are included in the data release.}
    \label{fig:output}
\end{figure}

The flowchart in Fig.~\ref{fig:output} shows an overview of the image-creation process, which will be described in more detail in this section.

\subsection{Roman Characteristics}

The photometric system and other characteristics of the telescope are based on \textit{Roman} Cycle 7 specifications. We briefly summarize useful characteristics here.  There are eight imaging filters of the WFI: R062, Z087, Y106, J129, H158, F184, K213, and W149 (a very wide filter). These filters will respectively be abbreviated in the text as $R$, $Z$, $Y$, $J$, $H$, $F$, $K$, and $W$. In $\mu$m, the central wavelengths of these filters are (respectively) 0.620, 0.869, 1.060, 1.293, 1.577, 1.842, 2.125, and 1.464, and the widths in $\mu$m of the transmission bands are 0.280, 0.217, 0.265, 0.323, 0.394, 0.317, 0.35, and 1.030, in the same order.\footnote{\url{https://roman.gsfc.nasa.gov/science/WFI_technical.html}} The spatial resolution of the imaging component of the WFI is $\sim 0.11\arcsec$ pixel$^{-1}$, and each HgCdTe Sensor Chip Array (SCA) image has a usable pixel grid of $4088 \times 4088$. The read noise is Gaussian with a standard deviation of 8.5 e$^-$.

A total of 18 SCAs make up the \textit{Roman} camera, arranged into a $6 \times 3$
array to generate an effective FoV of 0.281 sq. degrees. The fill factor across the field of view is $\sim 88$\%, owing to gaps between the SCAs. The simple pointing strategy we employ here (described below in Section \ref{sec:pointing}) does not include a dither strategy to deal with these chip gaps. Bright objects will further limit the usable field, as twelve diffraction spikes due to the \textit{Roman} secondary mirror support struts will be visible for most stars.

\subsection{Image Simulations from \citetalias{troxel_synthetic_2021}}

The simulation engine used for this work is the publicly available code developed for the \textit{Roman} HLIS, as described by \citetalias{troxel_synthetic_2021}.\footnote{\url{https://github.com/matroxel/roman_imsim}} Here, we explain the object creation (galaxies/stars) and image creation separately.

\subsubsection{Object Creation: Galaxies and Stars}\label{sec:object_creation}

The properties of our simulated galaxies are summarized in Table~\ref{tab:galaxy_properties}.
As shown there, we use three main inputs to generate astrophysical objects for the simulated images:
the spatial distribution of galaxies (from the Buzzard simulation), stellar population properties for the galaxies assigned to each position (from 3DHST catalogs), and foreground stars and dust (from  the Galaxia Milky Way (MW) simulation).

\begin{table*}
\caption{\label{tab:galaxy_properties} Properties of simulated objects in the Galaxy and Star input catalogs. All properties are discussed in Sec. 2.2.1.}
\begin{tabular}{lp{0.23\linewidth}lp{0.2\linewidth}}
Category & Property & Source & Reference \\
\hline
Location & RA, Dec & Buzzard & \citet{DeRose2019}\\
Redshift and Photometry & $z$, $Y$, $J$, $H$, $F$ mags, FWHM size & 3DHST SEDs & \citet{skelton_3d-hst_2014}\\
 & & $+$EAZY fits & \citet{brammer_eazy_2008} \\
Stellar Population & SFR, M & 3DHST catalogs & 
\citet{skelton_3d-hst_2014} \citet{whitaker_constraining_2014} \citet{momcheva_3d-hst_2016}  \\
Morphology & $a$, $b$, $\theta_{\rm rot}$, $f_{b}$, $f_{d}$, $f_{k}$ &  Random draws\\
MW properties & RA, Dec, and Y-F of MW stars, MW extinction & Galaxia & \citet{Sharma2011}\\
\hline

\end{tabular}
\end{table*}

The input galaxy distribution catalog is taken from one realization of the Buzzard simulation \citep{DeRose2019}, to introduce realistic spatial clustering. The Buzzard simulation provides the angular coordinates on the sky for $\sim 2.7$ million galaxies, across the full 5 sq. degrees of our simulated \textit{Roman} deep field.  Each position is then assigned a random set of galaxy properties drawn from the galaxy-properties catalog, which is based of the Cosmic Assembly Near-infrared Deep Extragalactic Legacy Survey (CANDELS).

We do not use the same CANDELS catalog as \citetalias{troxel_synthetic_2021}; instead, our galaxy properties are derived from a catalog of observed galaxies from the CANDELS/3DHST fields \citep{brammer_3d-hst_2012}.  This catalog is a subset of the v4.1 catalog produced by the 3DHST team \citep{momcheva_3d-hst_2016}, which contains $\sim 208,000$ objects.  
Our version of the catalog has been modified in two key ways.  First, we have selected the subset of objects that are identified as galaxies and have estimates for redshift, stellar mass, and star-formation rate (SFR), and also have a well-defined spectral energy distribution (SED) model. 
These selection cuts leave the sample with $\sim 112,000$ galaxies, which we use as an input library.
Second, we provide coefficients from  \cite{brammer_eazy_2008} that can be used for future studies to create simulated spectra for any galaxy in the catalog from template SEDs provided with \cite{brammer_eazy_2008} that will match the colors/properties of the CANDELS catalog. As we take galaxy colors directly from observed galaxies, host-galaxy extinction is already applied to the simulated galaxies.

\begin{figure*}
    \centering
    \includegraphics[draft=False,width=0.28\textwidth,angle=270]{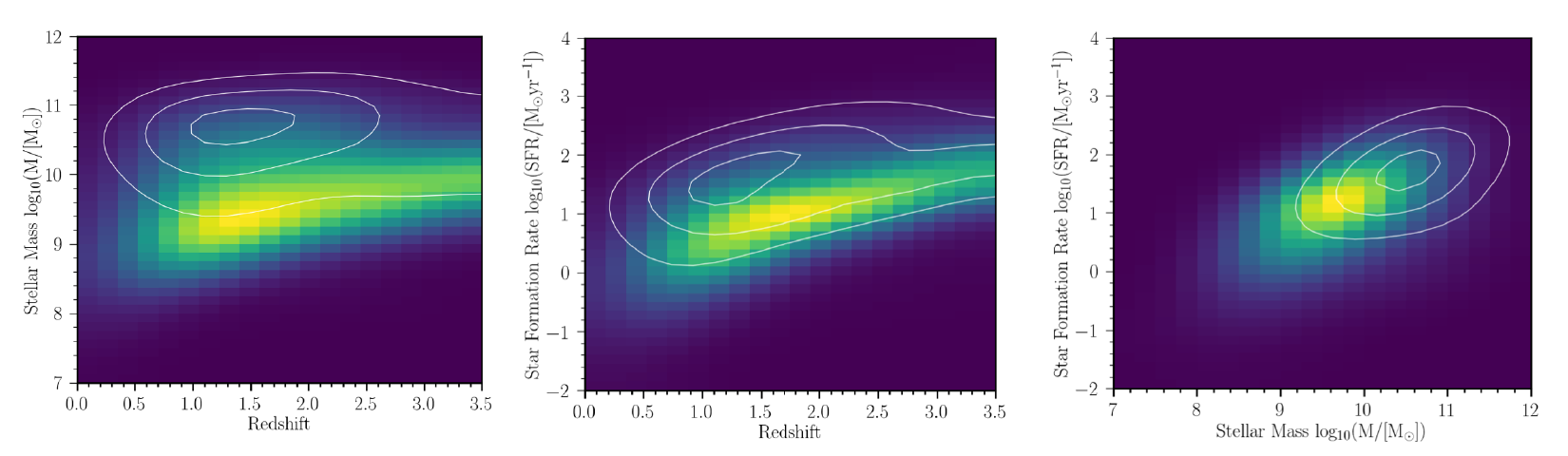}
    \caption{Stellar population properties of the galaxy catalog ($z<3.5$) used as input for \textit{Roman} High-Latitude Time-Domain Survey simulations.  Shading in each panel shows the distribution of the galaxy catalog with logarithmic scaling.  Overlaid contours show a representative sampling of SN~Ia host galaxies, drawn according to the specific-SFR-based model from \citet{AndersenHjorth}. Contours mark 30\%, 60\%, and 90\% levels.  Panel (a) presents the stellar mass vs. redshift distribution; (b) shows SFR vs. redshift; and (c) plots SFR against stellar mass.   The SN~Ia host-galaxy population is strongly skewed to higher mass and more actively star-forming galaxies at all redshifts.   }
    \label{fig:allgalaxies}
\end{figure*}

The redshift and mass estimates are described  by \citet{skelton_3d-hst_2014}. For the input redshift in our simulations, we adopt the spectroscopic redshift $z$ from the CANDELS catalog when available, and the photometric redshift otherwise. The mass estimates ($M$, in units of $\log_{10}\,(M/M_\odot)$) are derived from fits to the observed SEDs using the FAST code \citep{kriek_fast_2018}, and then corrected for overestimation of broad-band flux due to emission-line contamination, following \citet{whitaker_constraining_2014}.
We adopt the SFRs (in units of $(\log_{10}\, M_{\odot}) \, \textrm{yr}^{-1}$) derived from the combined ultraviolet (UV) and NIR flux measurements, as used by \citet{whitaker_constraining_2014}. Figure~\ref{fig:allgalaxies} shows the distribution of these galaxies across $z$, $M$, and SFR.
Finally, we generate photometric magnitudes for each galaxy by integrating the SED of each 3DHST galaxy, multiplied by the transmission functions of the \textit{Roman} bands used in the deep-field survey ($YJHF$).\footnote{See \url{https://roman.gsfc.nasa.gov/science/Roman_Reference_Information.html}} For this purpose we use SED fits generated by the 3DHST team with the EAZY code \citep{brammer_eazy_2008}.

As summarized in 
Table \ref{tab:galaxy_properties}, the following true object properties are assigned to each simulated galaxy: (1) redshift; (2) sky position in right ascension (RA) and declination (Dec) from the Buzzard catalog; (3) photometric properties (consistent $YJHF$
magnitudes); (4) base FWHM size as circular object;
%drawn from a random object in the photometric galaxy catalog;
(5) intrinsic shape components (defined using the distortion definition $|e| = (a^2-b^2)/(a^2+b^2), \, e_1 = |e| \cos{2\beta}, \, e_2 = |e| \sin{2\beta}$)
drawn from a Gaussian distribution of width 0.27 (truncated at
$\pm0.7$ to prevent impossible shapes); (6) random rotation angle; (7) ratio of fluxes in each
of the three galaxy components: (a) de Vaucouleurs bulge, $f_b$, (b) exponential disk, $f_d$, and (c) star-forming knots (at most 25\% of disk flux), $f_k$ --- created using GalSim's RandomKnots function; (8) stellar-population parameters (mass, SFR).

Each galaxy model is  built with GalSim from the truth parameters for the object in each \textit{Roman} bandpass.  The models are drawn in dynamically-sized square arrays of pixels (hereafter referred to as stamps). The size of each stamp is defined to include at least 99.5\% of the flux.

Following \citetalias{troxel_synthetic_2021}, we simulate the positions and magnitudes of input stars in the MW in \textit{Roman} passbands using the galaxy simulation Galaxia \citep{Sharma2011}.  Stars are simulated to mag 27 in the $V$ band, and the fluxes are converted to that measured with \textit{Roman} bandpasses using the stellar SED of $\alpha$~Lyrae derived from {\it Hubble Space Telescope (HST)} CALSPEC as packaged with GalSim. The star distribution has a mean stellar density of $\sim 2.5$~arcmin$^{-2}$. Extinction is added according to the Galaxia simulation using a formula based on the \citet{Schlegel1998} extinction maps.

\subsubsection{Image Creation}

\begin{figure}
\includegraphics[width=\columnwidth]{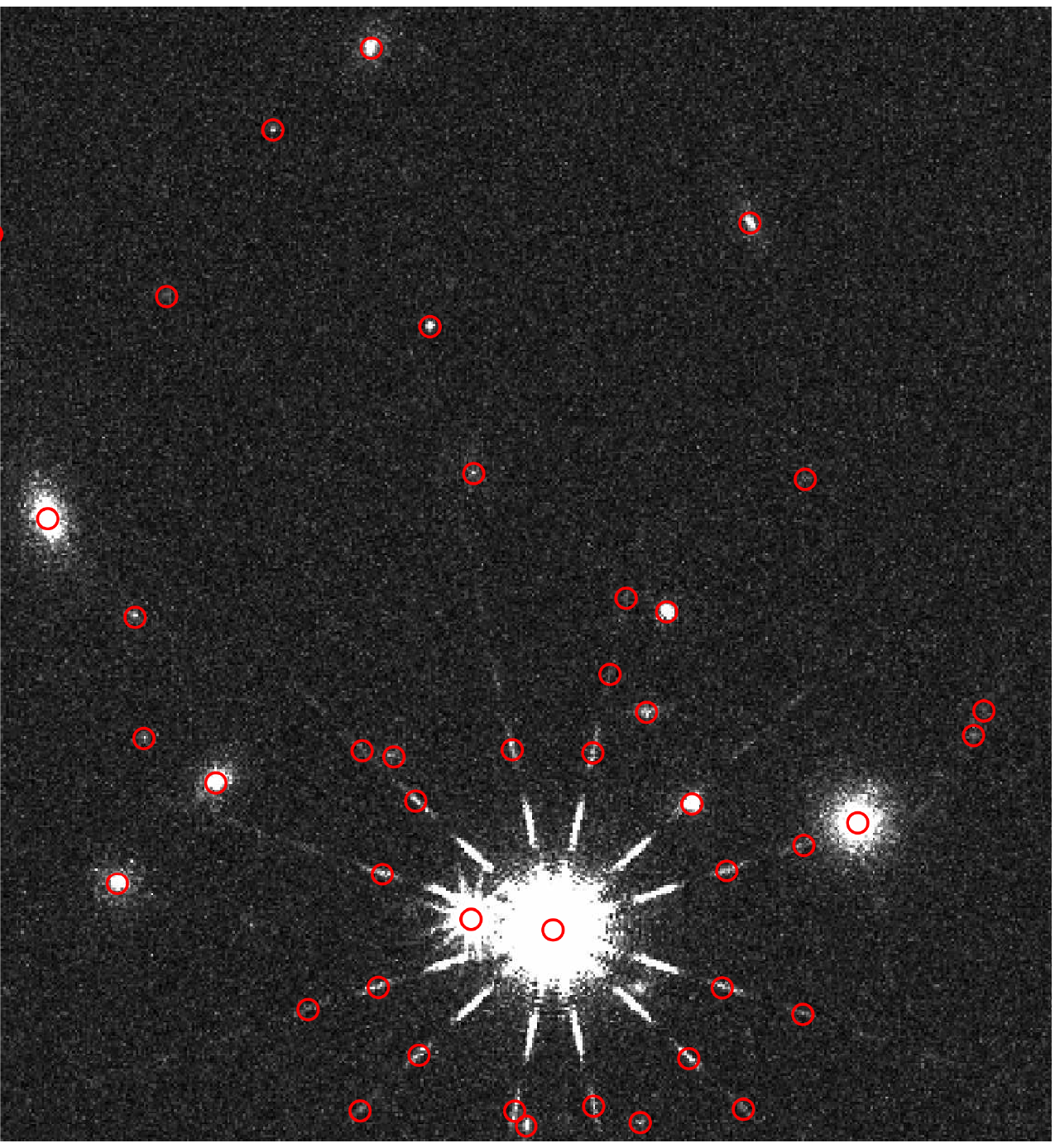}
\caption{An example cutout (0.75\arcmin $\times$ 0.75\arcmin) of a simulated image of one SCA that is 7.5\arcmin $\times$ 7.5\arcmin across.  The red circles show objects detected by Source Extractor that pass a S/N threshold.  Diffraction spikes of the stars can be seen.}
\label{fig:example_image}
\end{figure}

To create the image, the focal plane of the telescope is arranged at a given RA, Dec location with a given position angle. Each SCA has a fixed relationship with the focal plane, allowing its footprint to be determined from the information about the center point. Each image consists of a single SCA and is made up of $4088 \times 4088$ pixels. We identify the subset of galaxies and stars from the input catalogs that fall within that footprint, build a model for each object, and add them as stamps onto the SCA image. A cutout of one of the images is shown in Fig.~\ref{fig:example_image}. 

\citetalias{troxel_synthetic_2021} simulated images over 6 sq. degrees in the H158 filter but did not include transient sources and did not structure those simulations for time-domain analysis. In this analysis, we focus more on the time component in order to build a temporal series of images, and also inject additional \SNe{} as star-like point sources onto the images.  Still, the architecture of the simulations is the same, and we make top-level modifications to build the High-Latitude Time-Domain Survey.  

The images in the simulation are rendered using the GalSim software package \citep{Rowe2015}.  We use the galsim.roman module in GalSim release v2.3.2., which has a number of \textit{Roman}-specific implementation details \citep{galsim_roman}.\footnote{\url{https://github.com/GalSim-developers/GalSim}}

Each image is processed through a series of modifications, simulating the physical effects that would happen in the detector such as dark current, detector nonlinearity, and read noise. We do not include the effects of persistence between images. The details of these steps are given by \citetalias{troxel_synthetic_2021}. This module includes a high-resolution image of the \textit{Roman} spider pattern (i.e., the obscuration by the struts and camera in the pupil plane) for each filter and SCA that the point-spread function (PSF) model is generated from. Due to limitations in processing speed building the PSF model from this highly sampled pupil plane mask, we downgrade the original pupil plane mask image to a coarser binning (a downsampling of factor 8 for all objects but the brightest stars) to produce a fast, but still sufficiently accurate, approximation of the correct PSF model. For the brightest saturated stars, this produces visual artifacts in the simulated diffraction, and so for visual purposes we use a smaller down-sampling factor: a binning of 4 or 2 pixels is used for stars brighter than magnitude 15 and 12, respectively.

\subsection{Specific Survey Strategy}\label{sec:pointing}

\citetalias{hounsell_simulations_2018} proposed various strategies for the \textit{Roman} extragalactic survey. They created a realistic catalog-level simulation of the survey, for the history of observations as well as the library of detected SN light curves.  The inputs toward SNANA that we utilize here are the sequence of date and filter of each observation.

We choose to follow the Imaging:High-z strategy from \citetalias{hounsell_simulations_2018} with updates described in a new reference survey by \citet{RoseSurvey}. The strategy has a deep tier and a medium/wide tier, which cover 5 sq. degrees and 21 sq. degrees, respectively.  For this analysis, we focus solely on the deep tier owing to computational limitations.  The survey in the deep tier uses filters $YJHF$ (0.7--2.0 $\mu$m), and the exposure times are 300, 300, 300, and 900\,s for those filters, respectively.
The number of \SNe{} predicted by this strategy is $\sim 8,000$ using cutoffs for generation at $0.05<z<3.0$.  No SNR cuts are applied at this stage.

\begin{figure}
    \centering
    \includegraphics[width=\columnwidth]{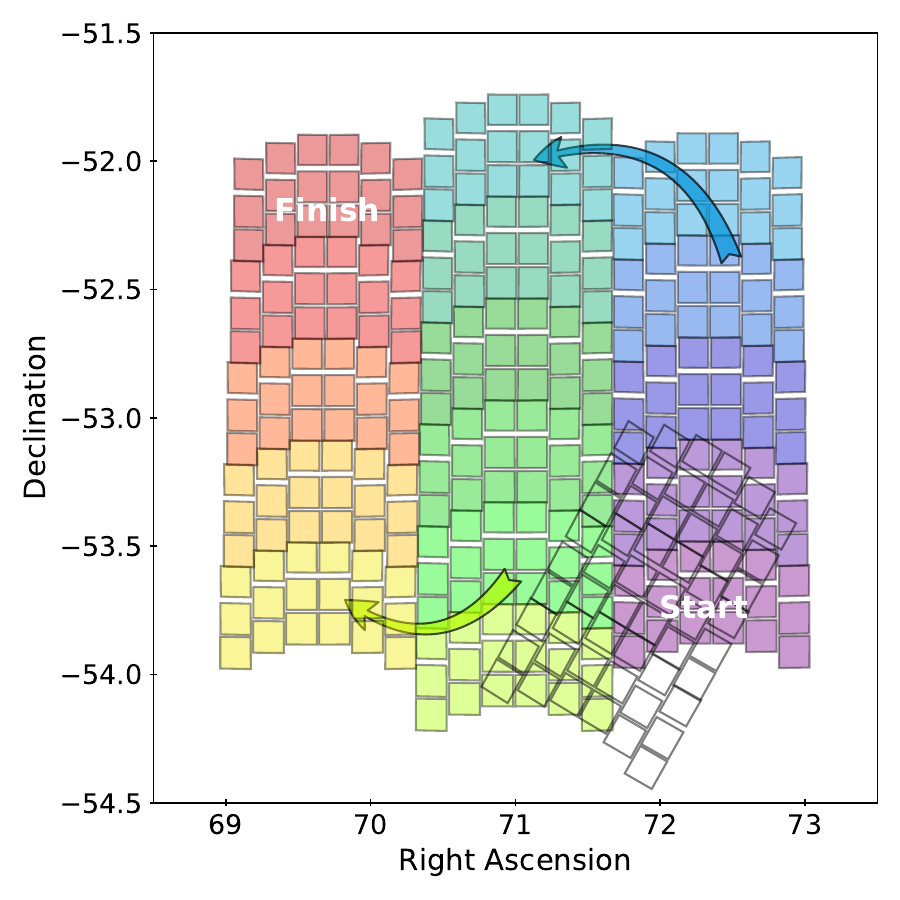}
    \caption{A portion of the pointing strategy. The pointing sequence for a single visit begins in the lower right (the purple WFI footprint, labeled ``Start") and proceeds up and down the columns as indicated by the arrows, ending in the upper left (red footprint labeled ``Finish").  The subsequent visit is rotated by $30^\circ$, and the first three pointings of that visit are shown as unfilled regions in the lower right.  An animation of this pointing strategy is available in the online version.}
    \label{fig:pointing_sub}
\end{figure}

For the creation of simulated images, we specify additional details of the observing strategy.  We choose the position to be in the Akari Deep Field South so that it is in a continuous viewing zone for longer observing seasons; the center position of the fields are chosen at 71$\degree$ RA, $-53\degree$ Dec.  The total survey spans 2 years with an observation cadence of 5 days, such that there are 146 visits to the SN field. \citetalias{hounsell_simulations_2018} did not need to specify the exact slewing strategy, as it was assumed all SNe could be discovered in the survey area. Here, we detail the chosen observation sequence for a 5 sq. degree field, which is illustrated in Fig.~\ref{fig:pointing_sub}.

\begin{itemize}
\item One orient of the telescope angle each month, and the angle is rotated by $30^\circ$ every 30 days.
\item For a given orient, there are 16 individual pointings that tile the field. The tiling is done such that there is one central row of 6 pointings and two parallel rows of 5 pointings.
\item Pointings are chosen to minimize gaps and the well-covered area is 5 sq. degrees, whereas the largest extent of any of the chips is 8 sq. degrees.
\item The sequence of pointings goes from southeast to northeast to north to south to southwest to northwest, using a ``snake'' pattern that minimizes distance between successive pointings (see animated version).
\item Within each column, the distance between pointings is $\sim 24$\arcmin, and the movement between columns is $\sim 49$\arcmin.
\item The pointing sequence is done in one filter, and once completed, the filter is switched and the sequence is done backward until each filter has been used.
\end{itemize}

For each ``snake'' pattern, the areas covered by SCAs do not overlap, so each SN~Ia will show up only in one SCA.

\begin{figure*}
    \centering
    \includegraphics[draft=False,width=0.4\textwidth,angle=270]{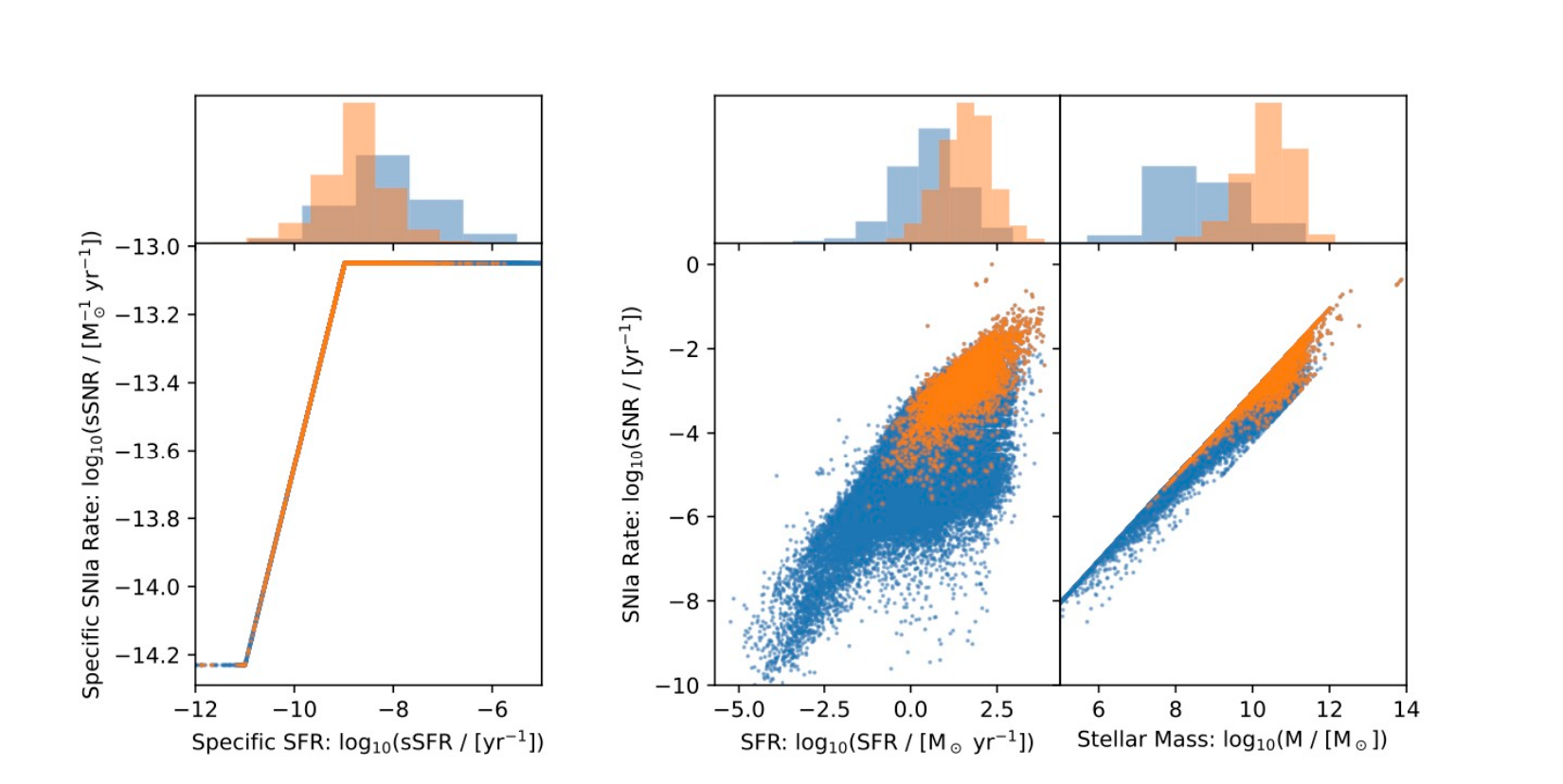}
    \caption{Application of a SN~Ia rate (SNR) model to the input galaxy catalog.  Each lower panel plots all 112,000 galaxies in blue, overlaid with a selection of 5000 ``SN~Ia host galaxies" in orange. Smaller panels above show normalized histograms for both populations.  The host-galaxy selection process follows the \citet{AndersenHjorth} specific SFR (sSFR)-based model to assign an SNR prediction to each galaxy in the catalog. That model is illustrated in the left pair of panels, plotting the specific SNR (sSNR) against the sSFR, with logarithmic scaling on both axes. The two pairs of panels on the right show the total SN~Ia rate for every galaxy, computed by multiplying the sSNR by the total stellar mass.  The middle panels plot this SNR against the SFR.  The far-right panels plot SNR against the stellar mass.
      }
    \label{fig:rodney_host}
\end{figure*}

\section{Including Supernovae in Images}\label{sec:supernovae}

Following \citetalias{hounsell_simulations_2018}, we use the same SN~Ia rates model \citep{Rodney2014,Graur2014}, light-curve model \citep{guy_supernova_2010}, SN~Ia population characteristics \citep{Scolnic16}, and area covered.  In order to assign \SNe{} to host galaxies, we follow the process outlined by \cite{popovic_assessment_2019}, defining properties of each simulated SN~Ia based on the properties of its simulated host galaxy. This allows us to propagate host-galaxy selection biases as well as specific correlations between SN~Ia properties and host-galaxy properties which can be used in a broader cosmological analysis of simulated data. Additionally, we vary the SN~Ia rates based on the host-galaxy properties, as discussed below.

\subsection{Matching Galaxy Catalogs to SN~Ia Catalogs and Injection of \SNe{} onto Images}

To project SN~Ia rates (SNRs) onto galaxy properties, we use the piecewise linear model from \citet{AndersenHjorth} which relates specific SFR (sSFR) to specific SNR (sSNR). We note that the SN~Ia rates prescribed here are relative at a given redshift, and the overall SN~Ia rates with redshift have no galaxy dependence as given in  \cite{Rodney2014,Graur2014}. The relations are shown in Figure \ref{fig:rodney_host}. Here, the piecewise function is used so that there are not unphysically high SNRs from galaxies that are outside of the range $-12<$ sSFR $<-8$, given as $\log_{10}(M_{star}\,M_{\rm galaxy}^{-1})\,\textrm{yr}^{-1}$, where we have well-calibrated SNR models.

Multiplying the sSNR by the total stellar mass, we derive the absolute SNR of each galaxy in the input catalog (see the far-right panels in Figure~\ref{fig:rodney_host}).  We then normalize these SNR values to get the relative probability of hosting a SN~Ia for each galaxy.

\begin{figure*}
    \centering
    \includegraphics[width=0.6\columnwidth]{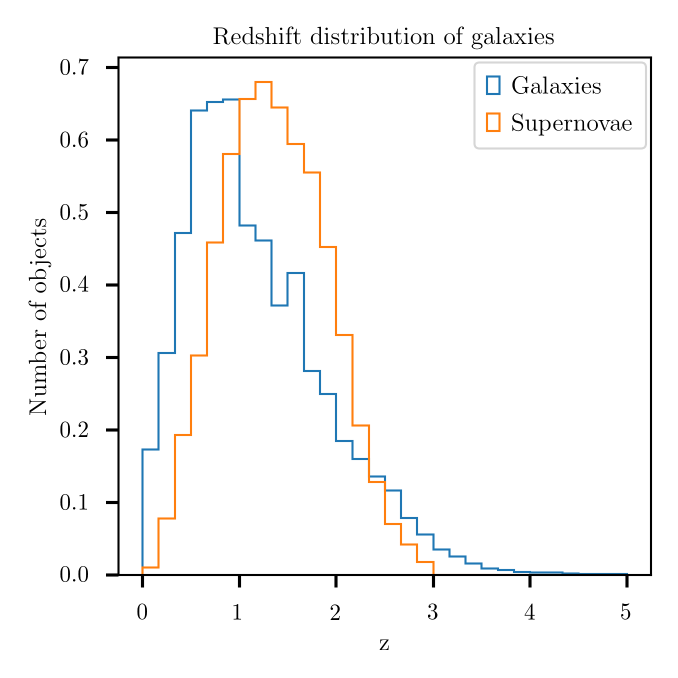}
    \includegraphics[width=0.6\columnwidth,angle=270]{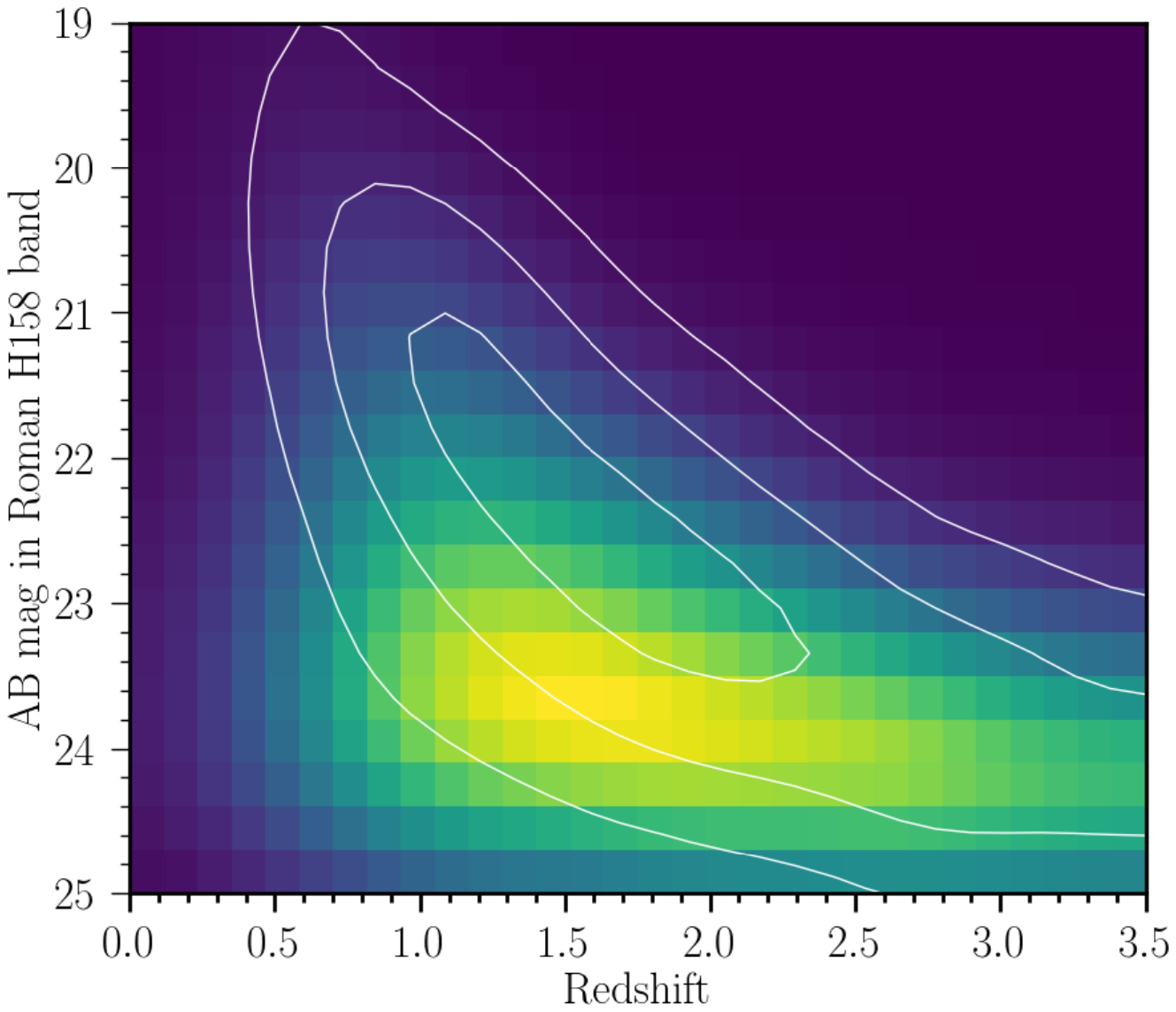}
    \caption{{\it Left:} The redshift distribution from the galaxy input catalog (blue) that is used as input for the image simulations and for the library of host galaxies of the \SNe{}.  The redshift distribution of the \SNe{} (orange) is capped at $z=3$ because limits on detection efficiency mean it is not useful to generate higher-redshift \SNe{}. 
    {\it Right:} Distribution of H158 magnitudes vs. redshift. Colored shading shows all galaxies in the input galaxy simulation catalog.  White contours show a subset of galaxies selected as SN~Ia hosts, based on the projected SN~Ia rate in each galaxy. Contours mark 30\%, 60\%, and 90\% levels. The host galaxies are brighter on average than the overall galaxy population at every redshift.}
    \label{fig:gal_mags}
\end{figure*}

The next step is to match simulated \SNe{} with appropriate hosts.  This begins with SNANA generating a set of simulated \SNe{} with redshifts in the range $0.06<z<2.99$.  For each SN~Ia we then identify all galaxies within $\Delta z \pm 0.01$ as the subset of potential host galaxies. The redshift distributions of galaxies and \SNe{} are shown in Figure \ref{fig:gal_mags}.  It is important to note that a SN~Ia is generated before assigning it to a host galaxy, which has the consequence that the redshift distribution of generated SNe will not match the redshift distribution of the galaxies.

A single galaxy is drawn from that subset, using the SNR probabilities as weights. The resulting distribution of stellar mass and SFR for host galaxies is plotted as contours in Figure \ref{fig:allgalaxies}. Selected host galaxies have on average higher stellar mass and SFR than galaxies overall. We also plot the magnitudes of the resulting host galaxies against redshift as contours in Figure \ref{fig:gal_mags}, which can be compared with the same relationship in all galaxies. The selection process using the weights causes host galaxies to be brighter on average than galaxies overall at any given redshift, an effect that can be seen directly in the images.

SNANA assigns an RA and Dec to the SN~Ia using the selected host galaxy's intensity profile.  The SN~Ia is placed within the region that encloses 99\% of the host-galaxy flux density, weighted by the galaxy's Sersic profile (so the SN~Ia distribution traces the host-galaxy light profile).  Finally, the output from the SNANA simulation is a library of light curves for \SNe{} at specific redshifts and positions based on the host galaxy, as well as peak brightness date along with various other properties relevant to \SNe{}.

We inject the \SNe{} as point sources in the exact way that stars are included in the images. The flux of each SN~Ia is defined per filter, in AB magnitudes. For our image simulations, we use the simulated magnitudes without noise added by SNANA because the image simulations contain their own noise. SNANA produces a truth file for each SN~Ia light curve which can be used in downstream analyses.

The time of observation defined by SNANA is not identical to the time of observation for any given simulated image, owing to the slewing sequence.  Therefore, we interpolate the SN~Ia light curve to determine the flux used when injecting each SN~Ia into an image. However, for this image set, the largest difference between the times created by SNANA and observation dates is 1/20\,day, so there should not be a large effect.

\begin{figure*}
    \centering
    \includegraphics[width=0.2\textwidth,angle=270]{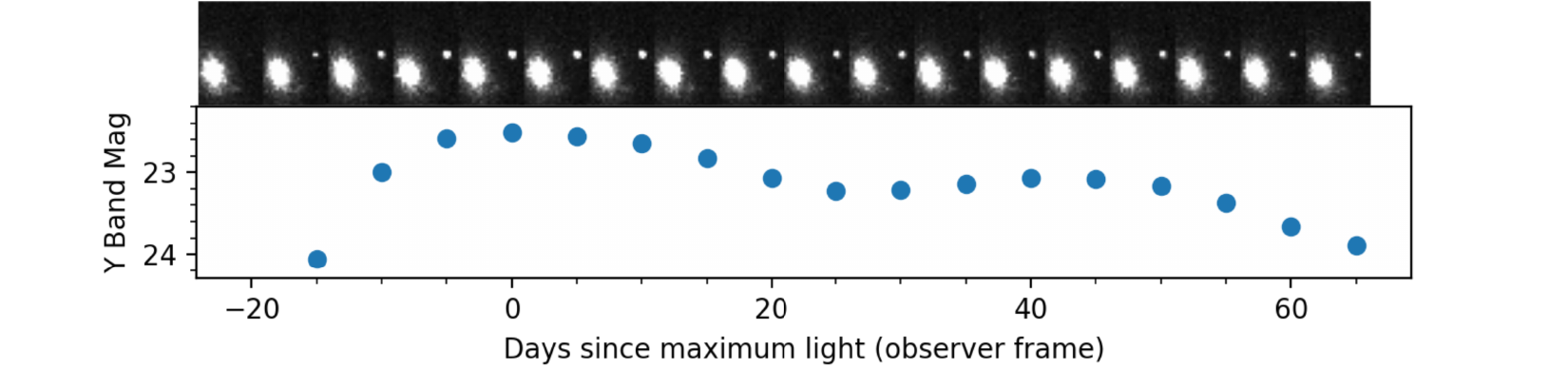}
    \caption{Simulated time series of a SN~Ia at redshift $z = 0.38$ in intervals of five days with corresponding light curve. Magnitudes are ideal values before noise and error are added. In each of the images, the SN~Ia is on the right with its host galaxy on the left. The images and light curve are aligned so that the points on the light curve match the image at the corresponding time.}
    \label{fig:lightcurve}
\end{figure*}

In Figure \ref{fig:lightcurve}, we give an example time series of a SN~Ia at redshift $z = 0.38$ injected near a galaxy in the $Y$ band.  The changing brightness of the SN~Ia can be seen to match the variation in brightness from the generated light curve.

\subsection{Image Data Products}
The image data, galaxy input catalogs, and SN~Ia light-curve catalogs are the three main products generated. These are publicly available.\footnote{\url{https://roman.ipac.caltech.edu/sims/SN_Survey_Image_sim.html}}

The images are created in a Flexible Image Transport System (FITS) format\footnote{\url{https://fits.gsfc.nasa.gov/}} and contain keyword information in their headers to make them compatible with {\it HST} MultiDrizzle software. The headers are created using GalSim's World Coordinate System (WCS) code and have WCS information in them. For each pointing, 18 separate image files are created, one for each SCA of the WFI.

While there are unchanging input catalogs for galaxy, star, and SN~Ia photometry, an additional truth file is created for each image containing every galaxy drawn on the image, containing each object's ID from the original input catalog, as well as their positions (both world coordinates and pixel positions). A similar truth file for stars and \SNe{} is created. We also include the magnitude drawn on the image in the truth file for \SNe{}.

\section{Analysis Examples}\label{sec:analysis}

In this section, we describe a series of analyses that we did using our simulated images and the tools we used for each analysis.

\subsection{Source Detection and S/N of Sources}\label{sec:Source Extractor}

We use Source Extractor \citep{SExtractor} to detect objects on the images. Source Extractor allows for the automated detection of objects from images and outputs estimated properties for each of those objects, including position, shape, signal-to-noise ratio (S/N), and other properties. It can be downloaded from its GitHub page.\footnote{\url{https://github.com/astromatic/SExtractor}}

Figure \ref{fig:example_image} shows one of our images, where we have highlighted objects detected by Source Extractor. We use default parameters for Source Extractor on our images except for the detection threshold (which is different between single images and coadds). 

\begin{figure}
\includegraphics[width=\columnwidth]{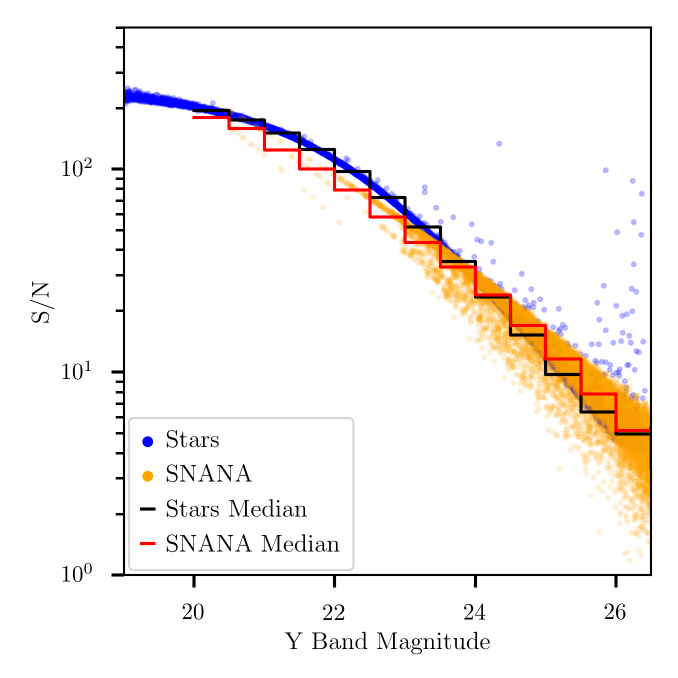}
\caption{A comparison of S/N for stars detected that were injected in $Y$ images (blue) and from our catalog-level simulations with SNANA (orange). Objects are also binned by $Y$-band magnitude, and the median S/N per bin is shown in black for stars and red from SNANA. For SNANA, the magnitudes are simulated magnitudes, while for stars, the magnitudes are measured magnitudes from Source Extractor. The small number of stars (in blue) with high SNR at faint magnitudes are due to blending with nearby sources.}
\label{fig:SNR}
\end{figure}

Using the images, we can also compare the S/N of point sources in the image to the S/N used in SNANA for catalog-level simulations. We use Source Extractor to obtain the S/N for stars in the images by taking the inverse of the root-mean-square (RMS) uncertainty in magnitude for each star using a 5\,pixel aperture diameter. We measure S/N for stars because they are more numerous than the \SNe{} and are not transient. However, because this does not include the contribution of galaxy noise, which is accounted for in the SNANA simulations, we add an approximate model as predicted by SNANA for a self-consistent comparison (on the level of 4 mmag). The results of the comparison are shown for $Y$-band images in Figure \ref{fig:SNR}. We find good agreement in S/N between the images and catalog-level simulation across the entire magnitude range.

\subsection{Stacking Images and SN~Ia Host-Galaxy Association}\label{sec:host_galaxy_associations}

\begin{figure*}
\includegraphics[width=0.25\textwidth,angle=270]{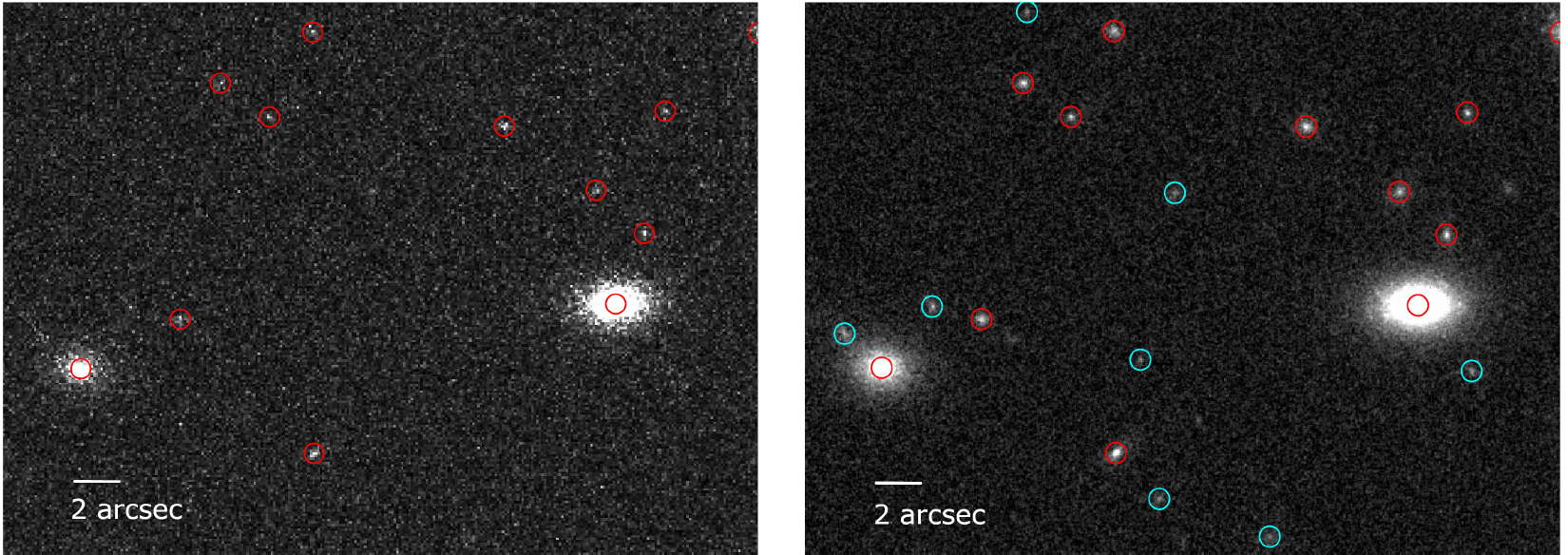}
\caption{{\it Left:} A single image with objects detected using Source Extractor circled in red. {\it Right:} The same area in the sky in a coadded image made by combining 64 images. The actual pixel depth is 10--15 images for most areas on the coadd. Objects found only in the coadded image are circled in cyan while objects found in the coadded image that were also found in the single image are circled in red.  There are no \SNe{} in this image.
}
\label{fig:coadd_comparison}
\end{figure*}

Several processing steps need to be done in order to reach the full potential use of the images taken by \textit{Roman}, including coadding images, discussed in this subsection, and difference imaging, discussed in the next. We have set up the process so that we can run the coaddition on an entire set of images, although there are currently still limitations in our mock coadd pipeline that limit how many images can be coadded together simultaneously.\footnote{Due to storage issues (how many images can be mapped pixel-by-pixel and stored in a 64-bit integer image bit-mask plane) as well as memory limitations when processing in parallel the coadd stage of the pipeline, we limit the number of overlapping images chosen to be coadded to 64. This is not a fundamental limit for Roman, but rather a limit of the current pipeline that was developed for this and related work using these image simulations.}

For these steps we use the sndrizpipe data-processing pipeline\footnote{\url{https://github.com/srodney/sndrizpipe}}, a Python implementation for \HST data processing designed for \HST SN surveys.  The sndrizpipe package was originally developed for detection of SNe in the CANDELS and FrontierSN {\it HST} surveys \cite[see][]{rodney_illuminating_2015,rodney_two_2015}.  It is built on the DrizzlePac toolkit, a package of software tools developed by the Space Telescope Science Institute (STScI) for processing of data from \HST and other observatories  \citep{Drizzle}.

In applying the sndrizpipe tools to the simulated \textit{Roman} images, we first select a ``WCS reference image'' --- an arbitrarily-selected image to define the world coordinate system (WCS) to which all other images will be registered.  Next, we use the TweakReg tool to construct a catalog of sources (stars and galaxies) in each individual-exposure image and match them with the WCS reference image's source catalog.   

We can then combine all the images using the MultiDrizzle algorithm implemented as the AstroDrizzle Python module.\footnote{\url{https://github.com/spacetelescope/drizzlepac}} We adopt an output pixel scale of 0.0575\arcsec and set the AstroDrizzle pixfrac parameter to 0.7. For now, we combine a limited set of images such that for the resulting coadded image, 10--15 single images contribute to most of its pixels. The current image simulations do not include cosmic rays, and so cosmic ray correction in the Drizzle process is turned off. Realistically simulating cosmic ray impacts would require simulating up-the-ramp fitting in the SCA measurements, which has not yet been implemented in these image simulations.

In the right side of Fig.~\ref{fig:coadd_comparison}, we show an example of a stacked \textit{Roman} image in $Y$ band. A single image for the same location is shown for comparison. The increase in depth can be clearly seen both in the noise and as objects detected (marked in cyan) which are not in the single image.

\begin{figure*}
\includegraphics[width=\columnwidth]{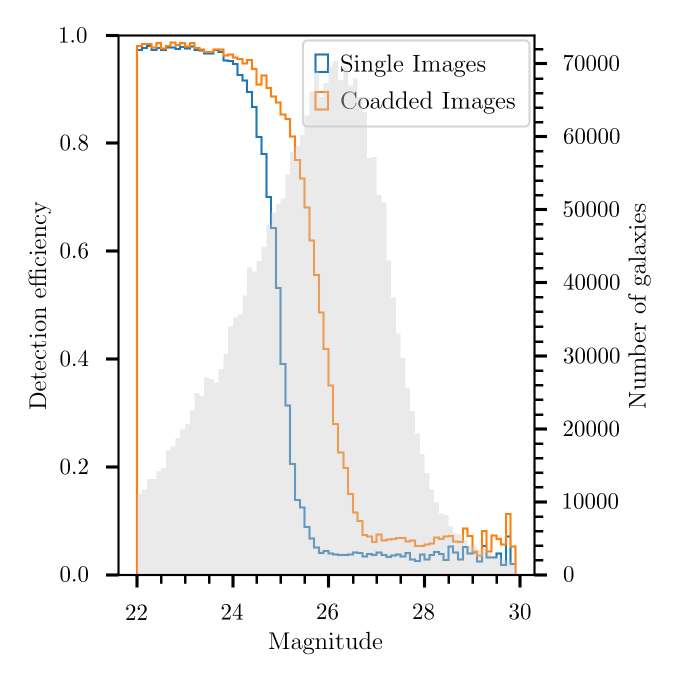}
% \caption{A comparison of galaxy detection efficiency in both single images (blue) and in coadded images (orange)}
% \label{fig:gal_efficiency}
\includegraphics[width=\columnwidth]{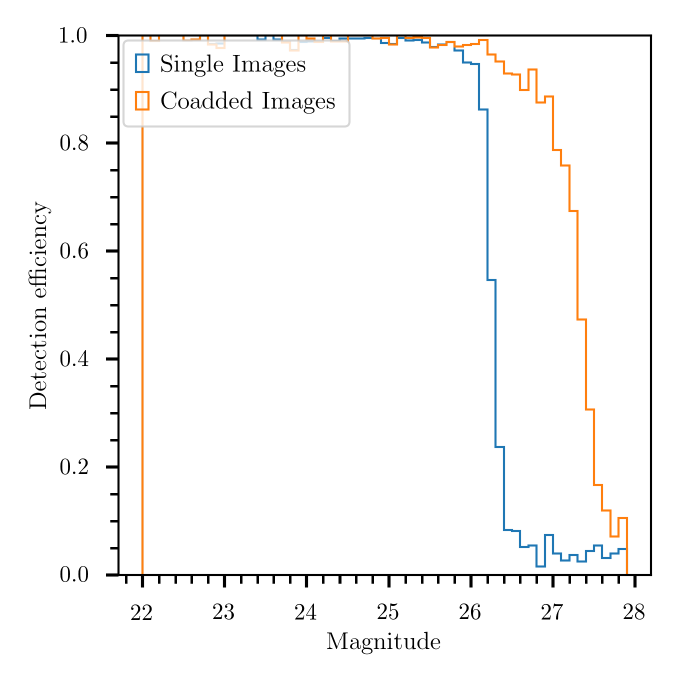}
\caption{{\it Left:} A comparison of galaxy detection efficiency in both single images (blue) and coadded images (orange). A histogram of the magnitudes of the galaxies is shown in filled gray. {\it Right:} A comparison of star detection efficiency in both single images (blue) and coadded images (orange). }
\label{fig:efficiency}
\end{figure*}

We performed a basic galaxy detection efficiency analysis with both single images and the stacked images described above in $Y$. A simple classifier is used, where we consider an object to be detected if Source Extractor detects an object within 0.3\arcsec of the truth value of its position. The results shown in the left side of Fig.~\ref{fig:efficiency} display the increase in depth obtained from stacking images. For single images, the 50\% recovery threshold is around 24.9\,mag. For coadded images, the 50\% recovery threshold is $\sim 26.0$\,mag, giving us about 1\,mag of depth over the single images. The CANDELS catalog goes above 30\,mag, so we expect some fraction of galaxies to remain undetected. Although we do not expect to detect faint (e.g., 29\,mag) objects, we do have a small percentage of false positives as a result of detections of unmasked features in the stellar diffraction spikes and of misidentification of overlapping detected bright galaxies near a faint background galaxy. This is due to the use of a simple matching algorithm by position on the sky, which is sufficient for the main results of the paper.

The same analysis can be performed on stars as well and is shown in the right side of Fig.~\ref{fig:efficiency}. Because stars and \SNe{} are both drawn as point sources on our images, these detection rates serve as useful proxies for potential SN~Ia detection rates. As point sources, stars are easier to detect than galaxies, resulting in 50\% recovery at 26.1\,mag and 27.3\,mag for single images and for coadded images, respectively. Note that unlike stars, SNe will not be detected in full-stack coadds because they will have faded away in some images. 

\begin{figure}
\centering
\includegraphics[width=\columnwidth]{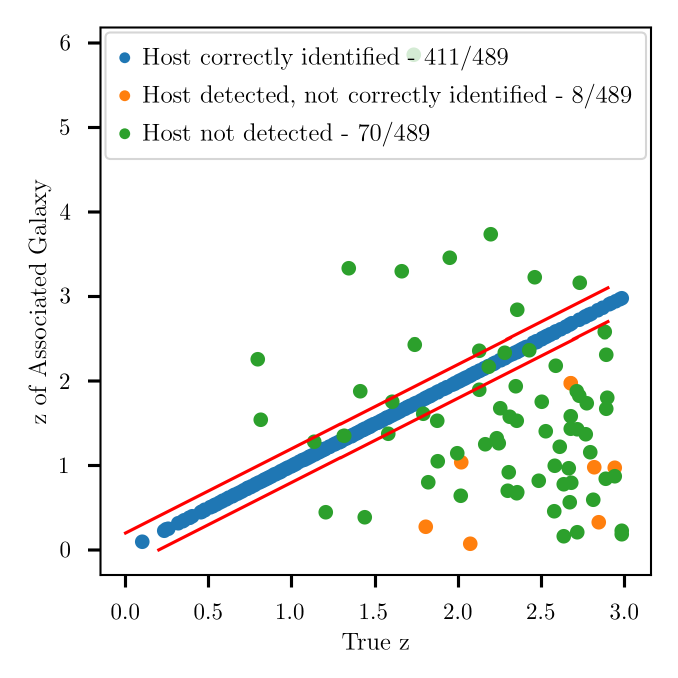}
\caption{A scatterplot of the true host-galaxy redshift and the redshift of the closest galaxy detected from each SN. Blue shows correct assignments, orange indicates when the host is detected but another galaxy is assigned as the host, and green signifies when the host is not detected and another galaxy is assigned as host.}
\label{fig:host_association_redshift}
\end{figure}

The \textit{Roman} High-Latitude Time-Domain Survey will rely on obtaining redshifts of the host galaxies to use \SNe\ on a Hubble diagram.  As discussed by \citet{Gupta16} and \citet{popovic_assessment_2019} with regards to DES and SDSS (respectively), there may be misassignment of the true host galaxy of a SN.  This can cause outliers on the Hubble diagram that cannot always be removed, and therefore cause a cosmological bias.  This may be especially problematic for \textit{Roman} owing to its large redshift range where fainter hosts will not be detected, as well as the possibility of blending \citep{Melchior2018} of galaxies over the images.

What the image set allows us that cannot be done at the catalog level is determining the galaxy detection limit, as some galaxies will be too faint to be detected with our algorithms. \SNe{} that belong to these galaxies may appear ``hostless" \citep{Sullivan06} or be assigned to a different galaxy.

Following \citet{popovic_assessment_2019}, we estimate the rate of misassociation using our coadded images.  \citet{popovic_assessment_2019} did this at the catalog level by assigning the host based on the separation and properties of nearby galaxies. We tested a similar procedure here by using the Source Extractor isophote parameters to define the elliptical effective radius with our injected artificial sources on real images.  Then we follow the directional light radius (DLR) method of \citet{Sullivan06} to assign host galaxies, which uses the distance from the galaxy to the center, as well as the orientation and shape of the galaxy. However, we found that simply using the distance between the SN~Ia and galaxy position gave better association results for our images.

The results of the host-galaxy association analysis are shown in Fig.~\ref{fig:host_association_redshift}. There were 489 host galaxies in one sample of images. 411 host galaxies are detected and correctly identified, as represented by the blue points. 8 host galaxies are detected but are not the closest galaxy, potentially leading to misassociation; these are shown as orange points. Furthermore, 70 host galaxies are not detected, which could also lead to misassociation. The redshift of the closest detected galaxy to the SN~Ia position is given in green. The set of \SNe{} (and thus host galaxies) span a wide redshift range. At lower redshifts, performance is much better, with only 2 out of 98 galaxies below $z = 1$ and 8 out of 234 galaxies below $z = 1.5$ not detected. 

A more sophisticated analysis could use the redshifts of the SN~Ia and of the suspected host galaxies to exclude obvious redshift mismatches between the SN~Ia and suspected host galaxy. If we assume that redshift differences greater than 0.2 would be detectable as outliers, indicated by the red lines in Fig.~\ref{fig:host_association_redshift}, we would correctly associate the 8 misassociated galaxies mentioned above. Finally, we note that this number of SNe without identified hosts will be underestimated slightly if SNe are located in galaxies that are too faint to be included in the CANDELS catalog used to generate galaxy properties; we note though that as shown in Fig.~\ref{fig:rodney_host}, we expect very few SNe to be in these types of galaxies, as the SFR and stellar mass are so low.

\subsection{Difference Imaging and Validating Simulated SN~Ia Detection Efficiencies}\label{sec:detection_efficiencies}

\begin{figure}
\centering
\includegraphics[width=\columnwidth]{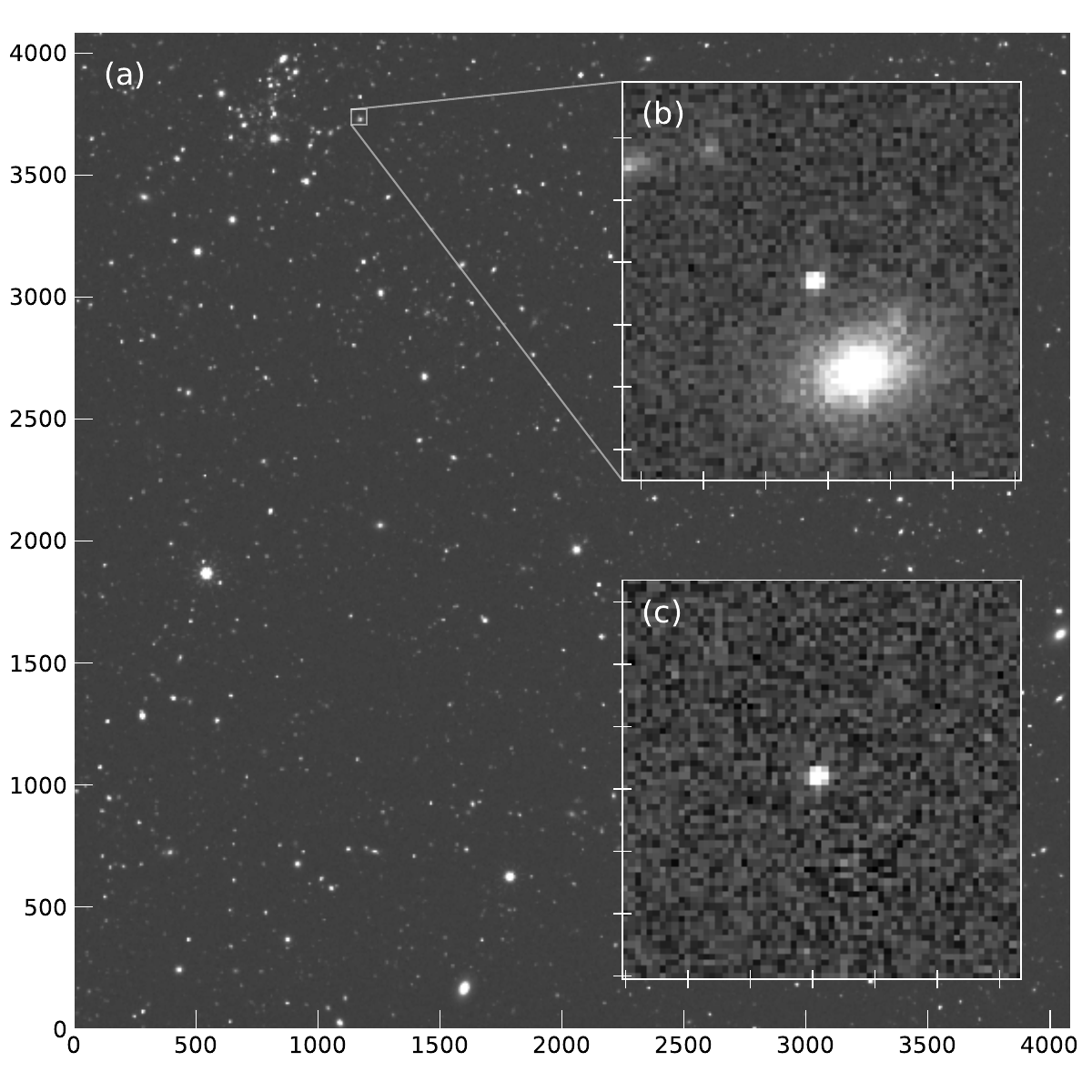}
\caption{A simulated $Y$-band image for one \textit{Roman} SCA. (a) The full frame of the simulated image, comprising $4088 \times 4088$ pixels, as indicated by the axis labels (roughly 7.5\arcmin on a side). (b) Zoom-in on a 10\arcsec $\times$ 10\arcsec region, centered on a bright SN~Ia with magnitude 22.59. (c) The difference image of the same 10\arcsec $\times$ 10\arcsec region, generated by subtracting a single-epoch template image taken after the SN~Ia has faded, with the same rotation and pointing center.  
}
\label{fig:subtraction}
\end{figure}

Given a template image, we can subtract it from each of the individual epochs.  As is typically done for space-based transient surveys, we do not include any convolution or other blurring of the image to match the PSF from epoch to epoch, because the PSF is presumed to be very stable across the time span of the imaging series (as discussed in T21). While there will be small variations in the PSF over time, mostly correlated with vibrational modes from movement of the telescope, we ignore these effects in this version of the simulation. The subtraction stage is therefore just a simple arithmetic operation at the end of the sndrizpipe pipeline.  The result is a series of difference images in which the galaxy light has mostly been removed, and we are left with isolated point sources (the simulated \SNe{}) and residual noise. Figure~\ref{fig:subtraction} shows a cutout of the observed image and difference image, centered on the location of a SN, which demonstrates that difference-imaging routines can work with these images.

As we do not have a full-scale pipeline to run this difference imaging on our set of simulated images, we work around that limitation by analyzing only \SNe{} which are isolated from their host galaxies. To ensure separation, we place a separation cut at 1.5 times the galaxy's semi-major axis and a host-galaxy brightness cut at 22\,mag.  We find that we can recover $50\%$ of the \SNe{} around a SN~Ia magnitude of 24.5\,mag and $0\%$ at 25\,mag.  While the results showcase the need for proper difference imaging, we are able to detect \SNe{} in our images, and the dropoff of efficiency for fainter magnitudes is expected owing to lower S/N.  For future SNANA catalog-level simulations, we can use the detection efficiency determined directly from the images, as there are likely features of the image analysis that cannot be fully modeled at the catalog level.

 \section{Conclusions}\label{sec:conclusions}
 
 This work is part of a larger ecosystem of simulations and analyses which will be used to optimize and prepare for the \textit{Roman} High-Latitude Time-Domain Survey.  One of the key elements of this work is that the catalog-level simulations are explicitly tied to the image-level simulations.  Follow-up analyses, like pixel-level simulations of prism spectroscopy with accurate SN~Ia host SEDs, can utilize the images created here.
 
 One main follow-up program is on improving the difference-imaging pipeline.  In recent surveys, typically two pipelines are written --- one for fast turn-around of transient detection and another with more accurate control of systematic uncertainties.  Examples of the former typically use image convolution and examples of the latter typically involve the scene-modelling approach.  One particularly interesting analysis would be to examine whether the ``surface-brightness anomaly," which shows unaccounted scatter in the uncertainty estimates of SN~Ia flux near bright galaxies, is seen in the \textit{Roman} image differences \citep{kessler_difference_imaging}.  A first use case would be to run the forward-modeling pipeline from \cite{Rubin21} on the images created here, as that study has indicated the anomaly may not be present for photometry of Roman images.
 
Another follow-up program is to create joint simulations of LSST and \textit{Roman} images.  There is currently focus on joint processing for weak-lensing studies, but a similar effort would benefit SN~Ia studies owing to increasing the wavelength and depth range.  Studies of this kind could be particularly fruitful for transient detection probabilities that are highly dependent on pixel-level effects or nonstandard image processing ---  e.g., transients near the cores of bright galaxies (includes lensed SNe, nuclear transients, quasars), as well as fast transients of various types, or very rare and faint transients like caustic-crossing stars and Population~III~SNe.  These future studies can leverage recent data releases with more NIR coverage of transients for better modeling of the transients (e.g. \citealp{Jones22}).

In summary, we have presented new simulated images for the \textit{Roman Space Telescope} that contain injected \SNe{}. These images could be used to look at effects such as detection efficiency near bright galaxies as well as to calibrate assumptions used for catalog-level simulations. We perform a variety of basic checks and analyses to demonstrate that \SNe{} were correctly drawn in the images and that the images can be useful for the aforementioned analyses.  As a product of this analysis, we release $1$ sq. degree of images over the full \textit{Roman} High-Latitude Time-Domain Survey and present these at \url{https://roman.ipac.caltech.edu/sims/SN_Survey_Image_sim.html}.  We encourage the community to use these images, and the code developed here can be further used for more simulations of different surveys/areas as well as studies of different transients.

\section*{Acknowledgements}

Much of this work is supported by NASA under Contract No. NNG17PX03C issued through the WFIRST Science Investigation Teams Programme. This work is based off of earlier work done by the Roman Cosmology with the High-Latitude Survey Science Investigation Team (https://www.roman-hls-cosmology.space/), which is supported by NASA Grant 15-WFIRST15-0008. D.S. is supported by DOE grant DE-SC0010007 and the David and Lucile Packard Foundation.  M.T. acknowledges support from NASA under JPL Contract Task 70-711320, “Maximizing Science Exploitation of Simulated Cosmological Survey Data Across Surveys.” A.V.F. is grateful for financial support from the Christopher R. Redlich Fund and many individual donors. This work is based on observations taken by the 3D-HST Treasury Program (GO 12177 and 12328) with the NASA/ESA HST, which is operated by the Association of Universities for Research in Astronomy, Inc., under NASA contract NAS5-26555. The material is based upon work supported by NASA under award number 80GSFC21M0002. Support for D.O. Jones was provided by NASA through the NASA Hubble Fellowship grant HF2-51462.001 awarded by the Space Telescope Science Institute, which is operated by the Association of Universities for Research in Astronomy, Inc., for NASA, under contract NAS5-26555.

\section*{Data Availability}

As stated in the paper, the simulated images are publicly released along with supporting files.\footnote{\url{https://roman.ipac.caltech.edu/sims/SN_Survey_Image_sim.html}} The code used to create the images is also publicly available,\footnote{\url{https://github.com/matroxel/roman_imsim}} as is the code used for the analysis.\footnote{\url{https://github.com/KevinXWang613/SNAnalysisCode}}

\bibliographystyle{mnras}
\bibliography{zotero_references.bib}

% Don't change these lines
\bsp	% typesetting comment
\label{lastpage}
\end{document}